\documentclass[12pt]{article}
\usepackage[hypertex]{hyperref}
\usepackage{amsmath,amssymb}
\usepackage{amsmath}
\numberwithin{equation}{section}

\newcommand{\Gammah}{\widehat{\Gamma}}
\newcommand{\Ah}{\widehat{A}}
\newcommand{\Psih}{\widehat{\Psi}}
\newcommand{\Phih}{\widehat{\Phi}}
\newcommand{\Fh}{\widehat{F}}
\newcommand{\epsilont}{\tilde{\epsilon}}
\newcommand{\Ncal}{{\cal N}}
\DeclareMathOperator*{\Tr}{{\rm Tr}}
\DeclareMathOperator*{\tr}{{\rm tr}}
\DeclareMathOperator*{\trp}{{\rm Tr}{}^{\prime}}
\newcommand{\del}{\partial}
\newcommand{\Gammat}{\widetilde{\Gamma}}

\def\s{\:\:\:\:}
\def\tepsilon{\widetilde\epsilon}
\def\tGamma{\widetilde\Gamma}

\begin{document}

%%%%%%%%%%%%%%%%%%%%%%%%%%%%%%%%%%%%%%%%%%%%
\thispagestyle{empty}
\begin{flushright}
OU-HET 707
\end{flushright}
\vskip3cm
\begin{center}
{\Large {\bf Towards the localization of SUSY gauge theory on a curved space}}
\vskip1.5cm
{\large 
{Koichi Nagasaki\footnote{nagasaki [at] het.phys.sci.osaka-u.ac.jp}
}
\,and\hspace{2mm} Satoshi Yamaguchi\footnote{yamaguch [at] het.phys.sci.osaka-u.ac.jp}
}
\vskip.5cm
{\it Department of Physics, Graduate School of Science, 
\\
Osaka University, Toyonaka, Osaka 560-0043, Japan}
\end{center}

%%%%%%%%%%%%%%%%%%%%%%%%%%%%%%%%%%%%%%%%%%%%
\vskip2cm
\begin{abstract}
We consider an $\Ncal=4$ supersymmetric gauge theory on a curved space.
We try to generalize Pestun's localization calculation on the four-sphere to a more general class of curved spaces.  We calculated the $Q$-exact term to localize the path-integral, and when it becomes positive definite, we obtain a configuration where the path-integral localizes. 
We also evaluate the super Yang-Mills action in this configuration.
\end{abstract}

%%%%%%%%%%%%%%%%%%%%%%%%%%%%%%%%%%%%%%%%%%

\newpage

\section{Introduction}

An interesting duality between 4-dimensional supersymmetric gauge theory and 2-dimensional conformal field theory was proposed by Alday, Gaiotto and Tachikawa(AGT) \cite{Alday:2009aq}.  In this duality, Nekrasov's instanton partition function \cite{Nekrasov:2002qd,Nekrasov:2003rj} in 4d gauge theory is equal to a conformal block in 2d conformal field theory. Moreover, when the equivariant parameters satisfy $\epsilon_1=\epsilon_2$, the partition function on $S^4$ obtained by Pestun \cite{Pestun:2007rz} is equal to a correlation function of 2d Liouville theory with $c=25$.

The localization method employed by Pestun \cite{Pestun:2007rz} is a very powerful technique to calculate exactly the partition function on $S^4$, Wilson loops and 'tHooft loops. These quantities were explored in \cite{Pestun:2009nn,Giombi:2009ek,Alday:2009fs,Drukker:2009id,Rey:2010ry,Nekrasov:2010ka,Passerini:2010pr,Gomis:2010kv,Gomis:2011pf}. This method was also applied to other backgrounds \cite{Dabholkar:2010uh,Nawata:2011un,Festuccia:2011ws}.  It is also useful in the 3-dimensional supersymmetric Chern-Simons theory \cite{Kapustin:2009kz} (See also \cite{Marino:2011nm} and references therein.).

One of the questions raised on AGT duality is what is the 4d counterpart of the 2d correlation function in the Liouville theory with central charges other than $c=25$.  One possibility is that it corresponds to the partition function of the gauge theory on another curved 4d space. Actually in 3-dimensions Hama, Hosomichi and Lee \cite{Hama:2011ea} have calculated the partition functions on squashed $S^3$ and have found that it is related to the Liouville theory with $c\ne 25$.
In 4-dimensions we will need to find some geometry on which the partition function includes Nekrasov's partition function with $\epsilon_1\ne \epsilon_2$.

Motivated by this problem we try to perform the localization procedure in a curved 4d spacetime in this paper.  We formulate the $\Ncal=4$ super Yang-Mills theory on a curved background which is locally obtained by a Weyl transformation from the flat background. We calculated the $Q$-exact term which will be used for the localization.  We found that the bosonic part of this term is not always positive definite. When it is positive definite we found the field configuration on which the path integral localizes. We also evaluate the super Yang-Mills action in this configuration.

The construction of this paper is as follows. In section \ref{sec:setup} we set up the notation and the SUSY gauge theory on the curved background.  In section \ref{sec:localization} we will show the explicit form of the $Q$-exact term and discuss its meaning. Section \ref{sec:discussion} is devoted to discussions.  In appendix \ref{app:calculation} the detailed calculation of the $Q$-exact term is given.

\section{Set up}
\label{sec:setup}
In this section we will set up the theory.  We mainly follow the notation of \cite{Pestun:2007rz}.

\subsection{The fields and the action}
The $\Ncal = 4$ theory can be obtained by the dimensional reduction from 10-dimensional super Yang-Mills theory.  In 4 dimensions, we have the gauge fields $A_{\mu},\ \mu=1,2,3,4$, the scalar fields $\Phi_{A},\ A=5,6,\dots,9,0$ and the spinor fields $\Psi$.

Let us start with the $\Ncal=4$ SYM on the flat 4-dimensional Euclidean space. The fields and the gamma matrices in flat space are denoted by the hatted notations $\Ah_{\mu}, \Psih, \Phih_{A}, \Gammah^{M}$.  The action of the SYM theory with the gauge coupling constant $g_{YM}$ in the 4-dimensional Euclidean space is given by
\begin{align}
 S=&\frac{1}{g_{YM}^2}\int d^4 x \trp\Bigg[
\frac{1}{4} \Fh_{\mu\nu}\Fh^{\mu\nu}+\frac12 D_{\mu}\Phih_{A}D^{\mu}\Phih^{A}
+\frac{1}{4}[\Phih_{A},\Phih_{B}][\Phih^{A},\Phih^{B}]\nonumber\\
&-\frac12 \Psih \Gammah^{\mu}\hat{D}_{\mu}\Psih
-\frac12 \Psih \Gammah^{A}[\Phih_A,\Psih]
\Bigg].\label{eq:flat-action}
\end{align}
Here $\trp[\cdot]:=-2\tr_{N}[\cdot]$ for SU$(N)$ gauge group\footnote{We put the factor $2$ here since it is rather common convention in the literature.}. For the other simple gauge groups, $\trp$ is defined such that it gives the positive definite inner product.

In this paper we will consider the curved space whose metric is given by $g_{\mu\nu}=e^{2\Omega(x)}\delta_{\mu\nu}$ with some Weyl factor $e^{2\Omega(x)}$.
The $\Ncal=4$ SYM action in this space is obtained by the following simple redefinition of the fields.
\begin{align}
 & \Ah_{\mu}=A_{\mu},\qquad \Psih=e^{\frac{3}{2} \Omega}\Psi,\qquad \Phih_{A}=e^{\Omega}\Phi_{A},\nonumber\\
 & \Gammah^{\mu}=e^{\Omega}\Gamma^{\mu},\qquad \Gammah^{A}=\Gamma^{A}.
\end{align}
The action \eqref{eq:flat-action} is rewritten by these new fields as
\begin{eqnarray}
 S&=&\frac{1}{g_{YM}^2}\int d^4 x \sqrt{g} \trp\Bigg[
	\frac{1}{4} g^{\mu\rho} g^{\nu\sigma} F_{\mu\nu} F_{\rho\sigma} 
	+\frac12 g^{\mu\nu} D_{\mu}\Phi_{A}D_{\nu}\Phi^{A} 
	+\frac{1}{12} R \Phi_{A}\Phi^{A}\nonumber\\
&&	\hspace{3cm}
	+\frac{1}{4}[\Phi_{A},\Phi_{B}][\Phi^{A},\Phi^{B}]
	-\frac{1}{2} \Psi \Gamma^{\mu}D_{\mu}\Psi	-\frac{1}{2} \Psi \Gamma^{A}[\Phi_A,\Psi]
	\Bigg],\label{eq:action}
\end{eqnarray}
where $R$ is the scalar curvature expressed in our space as
\begin{align}
 R=-6g^{\mu\nu}(\del_{\mu}\Omega \del_{\nu}\Omega+\del_{\mu}\del_{\nu}\Omega),
\label{R}
\end{align}
and $D_{\mu}\Psi$ is covariant in both the gauge transformation and the local rotation; it is defined as
\begin{align}
 D_{\mu}\Psi=\del_{\mu}\Psi+[A_{\mu},\Psi]+\frac14 \omega_{\mu}^{ab}\Gamma_{ab}\Psi.
\end{align}
Here $a,b$ are the labels of the local orthonormal basis, and $\omega_{\mu}^{ab}$ denotes the spin connection. If we put the explicit form of the Levi-Civita connection into this covariant derivative, the Dirac operator can be simplified as
\begin{align}
 &\Gamma^{\mu}D_{\mu}\Psi=\Gamma^{\mu}\del_{\mu}\Psi+\Gamma^{\mu}[A_{\mu},\Psi]+\frac{3}{2}\Gamma^{\mu}\del_{\mu}\Omega \Psi.
\end{align}
The covariant derivative for the scalar field is defined as the usual way $D_{\mu}\Phi_A=\del_{\mu}\Phi_A+[A_{\mu},\Phi_A]$.
\subsection{Supersymmetry}

The action \eqref{eq:action} is invariant under the superconformal transformation. 
\begin{subequations}
    \begin{eqnarray}
\delta A_{\mu}	&=&\epsilon \Gamma_{\mu} \Psi,\\
\delta \Phi_A	&=&\epsilon \Gamma_{A}\Psi,\\
\delta \Psi		&=&\frac12 F_{\mu\nu}\Gamma^{\mu\nu}\epsilon+D_{\mu}\Phi_{A}\Gamma^{\mu A}\epsilon +\frac12 [\Phi_A,\Phi_B]\Gamma^{AB}\epsilon
				+\frac12 \Gamma^{\mu A}\Phi_A \nabla_{\mu}\epsilon.\label{susy-spinor}
  \end{eqnarray}
 \end{subequations}
Here the superconformal transformation parameter $\epsilon$ is given by\footnote{These superconformal transformation parameters are bosonic in the convention in this paper.  Thus $\delta$ is fermionic.}
\begin{align}
 \epsilon=e^{\frac12 \Omega}\hat\epsilon
=e^{\frac12 \Omega}(\epsilon_{s}+x^{\mu}\Gammah_{\mu}\epsilon_c),
\end{align}
where $\epsilon_s$ and $\epsilon_c$ are constant spinors with 16 components. It is convenient to introduce the spinor denoted by $\epsilont$ defined by the following relation.
\begin{align}
& \nabla_{\mu}\epsilon=\Gammat_{\mu}\epsilont.
\end{align}
This $\epsilont$ can be explicitly written as
\begin{align}
 \epsilont=e^{-\Omega}\frac12 \del_{\nu}\Omega \delta^{\nu a}\Gamma_a \epsilon
 +e^{-\frac12 \Omega}\epsilon_c.
\end{align}
By using this notation, the superconformal transformation law for the spinor \eqref{susy-spinor} can be rewritten as
\begin{align}
 \delta \Psi=\frac12 F_{\mu\nu}\Gamma^{\mu\nu}\epsilon+D_{\mu}\Phi_{A}\Gamma^{\mu A}\epsilon +\frac12 [\Phi_A,\Phi_B]\Gamma^{AB}\epsilon
				-2\Phi_A\Gammat^{A} \epsilont.
\end{align}
%We suppose that the metric is derived from the flat metric by Weyl transformation.
%\begin{equation}
%g_{\mu\nu}=e^{2\Omega}\delta_{\mu\nu} dx^\mu dx^\nu
%\end{equation}
%where the parameter $\Omega$ is defined as a function of position.
%On general deformed surface, the $\Omega$ deformed factor is
%\begin{align}
% &e^{-2\Omega}=1+\frac{1}{16r^4}(u_1^2+u_2^2)^2+\frac{1}{2r^2}(\zeta_1 u_1^2+\zeta_2 u_2^2).
%\end{align}
%where the parameters $\zeta_1,\zeta_2$ are related to the equivariant parameter $\epsilon_1,\epsilon_2$ and we used the two parameters defined 
%\begin{equation}
%u_1:=x_1^2+x_2^2, \s u_2:=x_3^2+x_4^2.
%\end{equation}

\subsection{Wilson loop and off-shell supersymmetry}
The observable we will consider in this paper is the expectation value of a 1/2 BPS Wilson loop.  This Wilson loop has a circular shape in flat space.  This circle is parameterized with the parameter $\alpha$ as
\begin{align}
 x^{\mu}(\alpha)=(t \cos \alpha, t \sin \alpha, 0, 0),
\end{align}
where $t$ is the radius of the circle. The 1/2 BPS Wilson loop with this trajectory is defined by
\begin{align}
 W_{R}=\Tr_{R}P\exp\oint d\alpha \left(A_{\mu}\dot{x}^{\mu}+\Phi_0 \sqrt{g_{\mu\nu}\dot{x}^{\mu}\dot{x}^{\nu}}\right),\qquad \dot{x}^{\mu}:=\frac{dx^{\mu}}{d\alpha}.
\end{align}
This operator preserves half of the supersymmetry. In other words, this operator is invariant under the superconformal transformation with the parameter which satisfies the relation
\begin{align}
 \epsilon_c=\frac1t \Gamma^{012}\epsilon_s.\label{susy-wilson-loop}
\end{align}
This relation kills half of the supercharges and lets the others survive.

In order to use the localization method, we need a fermionic symmetry generator which is closed off-shell.  Let us fix the parameter $\epsilon$ such that it satisfies the condition \eqref{susy-wilson-loop} and
\begin{align}
 \Gamma^{1234}\epsilon_s=-\epsilon_s,\qquad \Gamma^{5678}\epsilon_s=+\epsilon_s,\qquad 
\epsilon_s \Gamma^0 \epsilon_s=\epsilon_s \Gamma^9 \epsilon_s=1.
\end{align}
It was found in \cite{Pestun:2007rz} that this supersymmetry can extended to off-shell by introducing auxiliary scalar fields $K_i,\ i=1,\dots,7$. The action is given by
\begin{align}
 S&=\frac{1}{g_{YM}^2}\int d^4 x \sqrt{g} \trp\Bigg[
\frac{1}{4} F_{MN} F^{MN}
+\frac{1}{12} R \Phi_{A}\Phi^{A}
-\frac12 \Psi \Gamma^{M}D_{M}\Psi
-\frac12 K_i K_i
\Bigg].\label{eq:off-shell-action}
\end{align}
Here we use labels $M,N=1,2,\dots,9,0$ and combine some of the terms.  This action \eqref{eq:off-shell-action} is equivalent to the original theory \eqref{eq:action} since by integrating out $K_i$ one will obtain the action \eqref{eq:action}.  The action \eqref{eq:off-shell-action} has an off-shell supersymmetry $Q$
\begin{subequations}
\begin{align}
& Q A_{M}=\epsilon \Gamma_{M}\Psi,\\
& Q \Psi=\frac12 F_{MN}\Gamma^{MN}\epsilon-2 \Phi_{A}\Gammat^{ A}\epsilont
 +K_i \nu_i,\label{QPsi}\\
& Q K_i=-\nu_i \Gamma^M D_{M} \Psi,
\end{align}
\end{subequations}

where $\nu_i,\ i=1,\dots,7$ are spinors related to $\epsilon$ and satisfy
\begin{subequations}
\begin{align}
& \epsilon \Gamma^{M}\nu_i=0,\label{nu1}\\
& \frac12 (\epsilon \Gamma_M \epsilon) \Gammat^{M}_{\alpha\beta}
=\nu^i_{\alpha}\nu^i_{\beta}+\epsilon_{\alpha}\epsilon_{\beta},\\
& \nu_i \Gamma^{M}\nu_j=\delta_{ij}\epsilon \Gamma^M \epsilon.
\end{align}
\end{subequations}

We use this supersymmetry to the localization in the next section.
\section{Localization}
\label{sec:localization}
We would like to introduce a method of localization, which allows us to evaluate the infinite-dimensional integral by computing the usual finite-dimensional integral.
%We calculate the value of action at critical configuration of $S^Q$.
%%%%%%%%%%%%%%%%%%%%%%%%%%%%%%
%\subsection{Localization}
The basic idea of the localization is as follows.  First let us deform the action by a $Q$-exact term.
\begin{equation}
Z_{\tau}= \int DAD\Psi\: e^{-S-\tau QV}.
\end{equation}
The derivative of $Z_{\tau}$ by $\tau$ turns out to be zero, and thus $Z_{\tau}$ is $\tau$ independent.  
Therefore we may evaluate it for whatever $\tau$ we want.  Usually it is convenient to take the limit $\tau\rightarrow \infty$.
\begin{equation}
Z=\int DAD\Psi\:  e^{-S}= \lim_{\tau \to \infty} \int DAD\Psi\: e^{-S-\tau QV}.
\end{equation}
As a result this integral is given by contributions from configurations which satisfy $QV=0$ and the 1-loop integral around them.

In this paper we choose $V$ as follows:
\begin{align}
 V= \int d^4x \sqrt{g}\trp(\overline{Q\Psi}\Psi)=:(\overline{Q\Psi}\:\Psi).
\end{align}
Here we use the simplified notation.
We want to calculate the bosonic part of $QV$ which is denoted by $S^Q_{bos}$.
\begin{align}
 S^{Q}_{bos}=QV|_{bosonic}=(\overline{Q\Psi}\:Q\Psi).
\end{align}
The detailed calculation is shown in appendix \ref{app:calculation}.
The result is given by
\begin{eqnarray}
S^Q_{bos}
&=& e^\Omega(F_{+}+w_{+}\Phi_9)^2 + (B-e^\Omega)(F_{-}+w_{-}\Phi_9)^2 -B\left( K_i+\frac{2}{B}\Phi_0\nu_i\tepsilon \right)^2\nonumber\\
&&	+ B\Big\{ D_m\Phi_i+\frac{1}{B}(\Phi_i\partial_m B+f_{mji}\Phi^j)\Big\}^2 \nonumber\\
&&	-B\left( D_\mu\Phi_0 +\Phi_0\frac{\partial_\mu B}{B}\right)^2
	+B\left( D_\mu\Phi_9+\Phi_9\frac{\partial_\mu B}{B}\right)^2 
	\nonumber\\
&&	+B[\Phi_0,\Phi_i][\Phi^0,\Phi^i]+B[\Phi_0,\Phi_9][\Phi^0,\Phi^9]+\frac{1}{2}B[\Phi_i,\Phi_j][\Phi^i,\Phi^j]\nonumber\\
&&	+\left[\left(\frac{1}{t^2}+\frac{1}{x^2}\right)64Y_{12}^2-\frac{1}{B}\partial_\mu B\partial^\mu B\right]\Phi_9^2\nonumber\\
&&	
	-\frac{1}{B}\left\{-3(\tepsilon \epsilon)^2+\frac{3}{2}\partial_\mu B\partial^\mu B +3B\cdot \tepsilon\tGamma^0\tepsilon\right\}\Phi_i^2.\label{result}
\end{eqnarray}
Here we use the notation
\begin{align}
& B:=\epsilon \Gamma^0 \epsilon=e^{\Omega}\left(1+\frac{x^2}{t^2}\right),\\
& f^{mij}:=\epsilont \Gamma^{0mij} \epsilon,\qquad m=1,2,3,4,9,\qquad i,j=5,6,7,8,
\end{align}

\begin{subequations}
\begin{align}
& w_+^{ab}:=-\frac{1}{e^{\Omega}}\epsilont^{L}\Gamma^{09ab}\epsilon^{L},\\
& w_-^{ab}:=-\frac{1}{B-e^{\Omega}}\epsilont^{R}\Gamma^{09ab}\epsilon^{R},
\end{align}
\end{subequations}

\begin{subequations}
\begin{align}
& W^{ab}:=e^{-\frac12\Omega}\frac12 \del_{\nu}\Omega \delta^{\nu a} x^{b},\\
& Y^{ab}:=W^{[ab]}|_{\text{self-dual}}.\qquad a,b=1,2,3,4.
\end{align}
\end{subequations}

$\epsilon^{L,R}$ and $\epsilont^{L,R}$ are also defined by

\begin{equation}
\begin{aligned}
 \epsilon=\epsilon^{L}+\epsilon^{R},\qquad \epsilont=\epsilont^{L}+\epsilont^{R},\\
 \Gamma^{1234}\epsilon^{L}=-\epsilon^{L},\qquad
 \Gamma^{1234}\epsilon^{R}=+\epsilon^{R},\\
 \Gammat^{1234}\epsilont^{L}=-\epsilont^{L},\qquad
 \Gammat^{1234}\epsilont^{R}=+\epsilont^{R}.
\end{aligned} 
\label{epsilonLR}
\end{equation}
This $S^Q_{bos}$ in eq. \eqref{result} is not always positive-definite because of the last two terms.  The positive-definiteness of $S^Q_{bos}$ requires the conditions
\begin{align}
 &\left[\left(\frac{1}{t^2}+\frac{1}{x^2}\right)64Y_{12}^2-\frac{1}{B}\partial_\mu B\partial^\mu B\right]\Phi_9^2\ge 0,\\
 &-\frac{1}{B}\left\{-3(\tepsilon \epsilon)^2+\frac{3}{2}\partial_\mu B\partial^\mu B +3B\cdot \tepsilon\tGamma^0\tepsilon\right\} \ge 0.
\end{align}

%%%%%%%%%%%%%%%%%%%%%%%%%%%%%%
%\subsection{The critical configuration of $S^Q_{bos}$}\label{CritiralConfig}
When both of these conditions are satisfied, the path integral has contributions from the zeros of $S^Q_{bos}$:
\begin{eqnarray}\label{CriticalPoint}
S^Q_{bos}=0 \longrightarrow
\left\{ \begin{array}{ll}
\Phi_0=\frac{a}{B},\s \Phi_i=\Phi_9=0,\s K_i=-\frac{2\nu_i\tepsilon}{B^2}a \\
\text{others}=0 \\
\end{array} \right. ,
\end{eqnarray}
where $a$ is a constant element of the gauge Lie algebra.
Substituting these values of fields, the action at this critical point is calculated as
\begin{eqnarray}
g_{YM}^2S &=& \int d^4x \sqrt{g}{\rm Tr'}\left[-\frac{1}{2}\partial_\mu\Phi_0\partial^\mu\Phi_0-\frac{R}{12}\Phi_0-\frac{1}{2}K_i^2\right]\nonumber\\
&=& ({\rm Tr'} a^2) \int d^4x \sqrt{g}\frac{1}{B^2}\left[2\frac{\tepsilon\tGamma^0\tepsilon}{B}-\frac{R}{12}\right].
\end{eqnarray} 
In this equation $\tepsilon\tGamma^0\tepsilon$ is evaluated as 
\begin{eqnarray}
\tepsilon\tGamma^0\tepsilon = -\frac{M^2+4Y^2}{t^2} - \frac{e^\Omega}{4}\partial_\mu\Omega\partial^\mu\Omega,
\end{eqnarray}
where $M$ is defined as $M:=\delta_{ab}W^{ab}+e^{-\frac12\Omega}$.
The scalar curvature of the space is given in eq. \eqref{R}. 
%\begin{equation}
%R=-6g^{\mu\nu}(\partial_\mu\Omega\partial_\nu\Omega+\partial_\mu\partial_\nu\Omega)
%\end{equation}
The Yang-Mills action $S$ at the configuration \eqref{CriticalPoint} becomes
\begin{equation}
g_{YM}^2S=-({\rm Tr'}a^2) \int d^4x \sqrt{g}\frac{e^{-2\Omega}}{B^2}
	\left[\frac{2(x^\mu\partial_\mu\Omega+1)}{t^2(1+\frac{x^2}{t^2})} -\frac{1}{2}\delta^{\mu\nu}\partial_\mu\partial_\nu\Omega\right].
\end{equation}

In order to evaluate the path-integral, we also have to consider the 1-loop determinants around these critical configurations. This problem will be discussed in a future work.
\section{Discussion}
\label{sec:discussion}
In this paper we discussed an $\Ncal =4$ supersymmetric gauge theory on a curved background and the localization. We calculated the $Q$-exact term and found the condition for positive-definiteness. We also found the critical configuration and the value of the super Yang-Mills action in this configuration.

The class of backgrounds we consider in this paper includes $AdS_2\times S^2$ and $AdS_4$.
\begin{align}
 &e^{-2\Omega}=1+\frac{1}{16r^4}(x^2)^2-\frac{1}{2r^2}(x_1^2+x_2^2-x_3^2-x_4^2),& (AdS_2\times S^2),\label{AdS2}\\
 &e^{-2\Omega}=\left(1-\frac{x^2}{4r^2}\right)^2,& (AdS_4),\label{AdS4}
\end{align}
where $r$ is a constant.
In particular $AdS_2\times S^2$ is the near horizon geometry of a 4-dimensional BPS black hole solution of $\Ncal=2$ supergravity.  The partition function on this background may be related to black hole statistical mechanics as pointed out in \cite{Dabholkar:2010uh}.
The black hole partition function is conjectured to be related to the absolute square value of the topological string partition function \cite{Ooguri:2004zv}. Furthermore the topological string partition function is Nekrasov's instanton partition function with $\epsilon_1=-\epsilon_2$ for a certain non-compact Calabi-Yau manifold. Therefore it is natural to expect that the partition function of the SUSY gauge theory on $AdS_2\times S^2$ is related to %$\epsilon_1=-\epsilon_2$ and 
the $c=1$ Liouville theory.
Unfortunately our $Q$-exact term is not positive-definite in these backgrounds. Thus we should find another good $Q$-exact term to investigate these backgrounds.

In the $S^4$ case, instantons localize at the North pole and anti-instantons localize at the South pole \cite{Pestun:2007rz}.  Let us guess from our result where the instantons and anti-instantons localize in other curved backgrounds.  Let us first consider $AdS_4$.
It is convenient to introduce a new coordinate $\rho$ by
\begin{align}
 \sqrt{x^2}=2r\frac{\sinh \rho}{\cosh\rho+1}.
\end{align}
Then the metric obtained by the Weyl rescaling by \eqref{AdS4} can be written as
\begin{equation}
ds_{AdS_4}^2=d\rho^2+\sinh^2\rho d\Omega_3^2,
\end{equation}
where $d\Omega_3^2$ is the metric of the unit three-sphere.
%where $\rho$ is a parameter parameterizing the radial direction in the $AdS$ space.
On this space the function $B(x)$ becomes
\begin{equation}
B(x):=e^\Omega\left(1+\frac{x^2}{t^2}\right) =e^\Omega\left(1+\frac{x^2}{4r^2}\right),\s t=2r,
\end{equation}
\begin{equation}
1+\frac{x^2}{4r^2}=\frac{2\cosh\rho}{\cosh\rho+1}.
\end{equation}
The location where instantons or anti-instantons localize is read off from the first and second terms of eq. \eqref{result}; instantons localize at $e^{\Omega}=0$ and anti-instantons localize at $B-e^{\Omega}=0$. In the case of $AdS_4$, $e^{\Omega}=\frac12(\cosh\rho+1)$ never becomes zero. On the other hand, $B-e^{\Omega}$ is calculated as
\begin{eqnarray}
B &=& \cosh\rho,\\
B-e^{\Omega} &=& \frac{1}{2}(\cosh\rho-1),
\end{eqnarray}
and it becomes zero when
\begin{equation}
B-e^{\Omega }=0 \longleftrightarrow \rho=0.
\end{equation}
Thus anti-instantons will localize at the center of $AdS_4$.

Second, $AdS_2\times S^2$ obtained by eq. \eqref{AdS2} is formulated by the following metric with the coordinates $(\rho,\theta,\psi,\phi)$
\begin{equation}\label{DefExp2OmegaAdS2S2}
ds_{AdS_2\times S^2}^2=r^2(d\rho^2 +\sinh^2\rho d\psi^2+d\theta^2 +\sin^2\theta d\phi^2).
\end{equation}
The coordinates $(\rho,\theta,\psi,\phi)$ are related to the coordinates $x_{\mu}$ by
\begin{equation}
\begin{aligned}
 &\sin \theta=4r\frac{\sqrt{x_3^2+x_4^2}}{\sqrt{(x^2-4r^2)^2+16r^2(x_3^2+x_4^2)}},\\
 &\sinh \rho =4r\frac{\sqrt{x_1^2+x_2^2}}{\sqrt{(x^2-4r^2)^2+16r^2(x_3^2+x_4^2)}},\\
 &\tan \psi=\frac{x_2}{x_1},\qquad \tan \phi=\frac{x_4}{x_3}.
\end{aligned} 
\end{equation}
The function $B$ in this background can be written as
\begin{equation}
B=e^{\Omega}\left(1+\frac{x^2}{4r^2}\right)=\frac{\cosh\rho-\cos\theta}{2}\frac{2\cosh\rho}{\cosh\rho-\cos\theta}=\cosh\rho.
\end{equation}
The instanton configuration localizes at $e^{\Omega}=0$. 
\begin{align}
 e^{\Omega}=\frac12(\cosh \rho -\cos \theta)=0.
\end{align}
This is the center of $AdS_2$ and the North pole of $S^2$.
On the other hand, the anti-instanton configuration localizes at
\begin{equation}
B-e^{\Omega}=\frac{\cosh\rho+\cos\theta}{2}=0.
\end{equation}
This is the center of $AdS_2$ and the South pole of $S^2$.
This picture is quite similar to \cite{Gaiotto:2006ns,Beasley:2006us}

%Let us summarize the discussion of the location of instantons. We concluded that in general Instanton in $AdS$ space localizes at the center of it and Instanton in an arbitrary dimension sphere localizes at two points, corresponding to two poles of the globe.
%%%%%%%%%%%%%%%%%%%%%%%%%%%%%%%%%%%
\subsection*{Acknowledgments}

We would like to thank Etsuko Itou, Takahiro Kubota and Takahiro Nishinaka for many illuminating discussions, important comments and suggestions. We also thank Wade Naylor for a careful reading of this manuscript and useful comments.
S.Y. was supported in part by KAKENHI 22740165.
%%%%%%%%%%%%%%%%%%%%%%%%%%%%%%%%%%5
\appendix
\section{Detailed calculation}
\label{app:calculation}
The purpose of this appendix is to calculate $S^Q_{bos}$, namely, the derivation of $S^Q_{bos}$ and rewriting it into the complete square form.
%This analysis shows that the location of instanton and each fields configuration are localized in some region of the manifold.

\subsection{Derivation of $S^Q_{bos}$}
We develop the deformation term of the action $S^Q$ to use the localization method. Especially the bosonic term $S^Q_{bos}$ is an important factor in determining the configuration of the fields.

The form of the functional $V$ can be written in the same way as \cite{Pestun:2007rz}:
\begin{eqnarray}
V&=&(\overline{Q\Psi}   \:\:\Psi) 
	\nonumber\\
&=&   \int d^4x \: \sqrt{g}\trp 
	\Big(\frac{1}{2}\epsilon\Gamma^0 \Gamma^{NM} F_{MN}
	  +2\Phi_A \tilde\epsilon\tilde\Gamma^0 \Gamma^A
	  -K_i \nu_i \Gamma^0\Big)\Psi.
\end{eqnarray}
In the second line the integral and trace are written explicitly.
For the sake of simplicity, we omit these symbols.

The bosonic part of $S^Q$ is given by
\begin{equation}
S^Q_{bos}:=QV|_{bos}=(\overline{Q\Psi}   \:\:Q\Psi) ,
\end{equation}
where $Q\Psi$ is given in eq.\eqref{QPsi} 
\begin{align}\label{QV}
Q\Psi &=
	 \frac{1}{2}F_{MN} \Gamma^{MN} \epsilon
		-2\Phi_A \tilde\Gamma^A \tilde\epsilon +K_i \nu_i,
\end{align}
and $\overline{Q\Psi}$ is defined by
\begin{subequations}\label{QV2}
\begin{align}
\overline{Q\Psi} &:=
\frac{1}{2}F_{MN} \tilde\Gamma^{MN} \Gamma^0 \epsilon
	+2\Phi_A \Gamma^A \tilde\Gamma^0 \tilde\epsilon
	-K_i\Gamma^0\nu_i,\\
\overline{Q\Psi}^T &=
	\frac{1}{2}F_{MN}\epsilon \Gamma^0 \Gamma^{NM}
	+2\Phi_A \tilde\epsilon \tilde\Gamma^0 \Gamma^A
	-K_i \nu_i \Gamma^0.	
\end{align}
\end{subequations}

Here the parameter $\epsilon$ satisfies the following conditions
\begin{subequations}
\begin{align}
 &\epsilon=e^{\frac12 \Omega}(\epsilon_s+x^a \Gammat_a \epsilon_c),\\
 &\epsilont=e^{-\frac12\Omega}\left(
\frac12\del_{\nu}\Omega\delta^{\nu a}\Gamma_a (\epsilon_s+x^b\Gammat_b \epsilon_c)
+\epsilon_c
\right).
\end{align}
\end{subequations}

\begin{subequations}
\begin{align}
 &\Gamma^{1234}\epsilon_s=-\epsilon_c,\qquad \Gamma^{5678}\epsilon_s=+\epsilon_s,\\
 &\Gammat^{1234}\epsilon_c=-\epsilon_c,\qquad \Gammat^{5678}\epsilon_c=+\epsilon_c,\\
 &\epsilon_c=\frac{1}{t}\Gamma^{012}\epsilon_s.
\end{align}
\end{subequations}
Substituting eqs.\eqref{QV},\eqref{QV2}, $S^Q_{bos}$ is obtained as summation of the following six terms.

\begin{align}
S_{FF} &= \frac{1}{4} F_{MN} F_{PQ}
	\epsilon \Gamma^0\Gamma^{NM}\Gamma^{PQ} \epsilon
=
{
	\frac{1}{2}BF_{MN}F^{MN}-\frac{1}{4}F_{MN}F_{PQ}\epsilon\Gamma^0\Gamma^{MNPQ}\epsilon
	}\label{SFF},\\
%%%%%%%%%%%%%%%%%%%%%%%%%%%%%%%%%%%%%%%%
S_{\Phi\Phi}&= -4\Phi_A\Phi_B \tepsilon\tGamma^0\Gamma^A\tGamma^B\tepsilon
={ 
	-4\Phi_A\Phi^A\tepsilon\tGamma^0\tepsilon
	},\\
%%%%%%%%%%%%%%%%%%%%%%%%%%%%%%%%%%%%%%%%
S_{KK}&=-K_ik_j\nu_i\Gamma^0\nu_j
={
-BK_iK_i
},\\
%%%%%%%%%%%%%%%%%%%%%%%%%%%%%%%%%%%%%%%%
S_{F\Phi}&= -f_{MN}\Phi_A\epsilon\Gamma^0\Gamma^{NM}\tGamma^A\tepsilon 
	+\Phi_AF_{PQ}\tepsilon\tGamma^0\Gamma^A\Gamma^{PQ}\epsilon\nonumber\\
&= {
	F_{MN}\Phi_A \tepsilon
	\Big(
           \tGamma^A \tGamma^{NM} \Gamma^0 + \tGamma^0 \Gamma^A \Gamma^{MN}
	\Big)\epsilon
	},\label{SFPhi}\\
%%%%%%%%%%%%%%%%%%%%%%%%%%%%%%%%%%%%%%%%
S_{\Phi K}&= 2\Phi_AK_j\tepsilon\tGamma^0\Gamma^A\nu_j 
	+K_j2\Phi_A\nu_i\Gamma^0\tGamma^A\tepsilon\nonumber\\
&= {
4\Phi^0 K_i \tepsilon \nu_i
},\\
%%%%%%%%%%%%%%%%%%%%%%%%%%%%%%%%%%%%%%%%
S_{FK}&= \frac{1}{2}F_{MN}K_j\epsilon\Gamma^0\Gamma^{NM}\nu_j
	-K_j\frac{1}{2}F_{MN}\nu_j\Gamma^0\Gamma^{MN}\epsilon\nonumber\\
&= \frac{1}{2}F_{MN}K_j\epsilon
	\Big(
	-\Gamma^0\Gamma^{MN}+\tGamma^{MN}\Gamma^0
	\Big)\nu_j\nonumber\\
&= -2F_{MN}g^{M0}K_j\epsilon\Gamma^N\nu_j\nonumber\\
&= 0.
\end{align}
In the last equality we use the formula \eqref{nu1}.  In these equations we use the notation
\begin{align*}
 B(x):=\epsilon\Gamma^0 \epsilon=e^{\Omega(x)}\left(1+\frac{x^2}{t^2}\right).
\end{align*}
If the spacetime is $S^4$ as in \cite{Pestun:2007rz}, $B(x)=1$, otherwise in general $B(x)$ depends on $x$.
$S^Q_{bos}$ is sum of these terms
\begin{equation}
S^Q_{bos}=S_{FF}+S_{\Phi\Phi}+S_{KK}+S_{F\Phi}+S_{\Phi K}+S_{FK}.
\end{equation}
Let us calculate the terms \eqref{SFF} and \eqref{SFPhi} in detail.

%%%%%%%%%%%%%%%%%
\subsubsection{Calculation of $S_{FF}$}
The indices are classified into classes: four-dimensional space-time, $\mu,\nu,\cdots=1,2,3,4$, extra-dimensional coordinates, $i,j,\cdots=5,6,7,8$ and $0,9$.
The possible combinations of the indices are listed as
\begin{eqnarray}\label{BAAIWAKE}
\{M,N,P,Q\}&=& \{\mu,\nu,\rho,\sigma\},\{9,\mu,\nu,\rho\},\{0,\mu,\nu,\rho\},\nonumber\\
		&&	\{i,j,k,l\},\{9,\mu,i,j\},\{\mu,\nu,i,j\}.
\end{eqnarray}
The results of each terms are shown here.
\begin{enumerate}
\def\theenumi{{ \roman{enumi})}}
 \item $\{M,N,P,Q\}=\{\mu,\nu,\rho,\sigma\}$
\begin{eqnarray}
-\frac{1}{4}F_{MN}F_{PQ}\epsilon\Gamma^0\Gamma^{MNPQ}\epsilon
&=& -\frac{1}{4}F_{MN}F_{PQ}\epsilon\Gamma^0\Gamma^{MNPQ}\epsilon\nonumber\\
&=& -\frac{1}{2}(B-2e^\Omega)F_{\mu\nu}*F^{\mu\nu},
\end{eqnarray}

\item $\{M,N,P,Q\}=\{9,\mu,\nu,\rho\}$
\begin{equation}
4\times \left(-\frac{1}{4}\right)F_{9\mu}F_{\nu\rho}\epsilon\Gamma^0\tGamma^{9\mu\nu\rho}\epsilon
=4\tepsilon\Gamma^{90}\Gamma^{\mu\nu}\epsilon\Phi_9F_{\mu\nu},
\end{equation}

\item $\{M,N,P,Q\}=\{0,\mu,\nu,\rho\}$
\begin{equation}
4\times \left(-\frac{1}{4}\right)F_{0\mu}F_{\nu\rho}\epsilon\Gamma^0\tGamma^{0\mu\nu\rho}\epsilon
=F_{0\mu}F_{\nu\rho}\epsilon\Gamma^{\mu\nu\rho}\epsilon=0,
\end{equation}

\item $\{M,N,P,Q\}=\{i,j,k,l\}$
\begin{equation}
-\frac{1}{4}F_{ij}F_{kl}\epsilon\Gamma^0\Gamma^{ijkl}\epsilon
=-\frac{1}{4}F_{ij}\epsilon^{ijkl}F_{kl}\epsilon\Gamma^0\Gamma^{5678}\epsilon
=0,
\end{equation}

\item $\{M,N,P,Q\}=\{9,\mu,i,j\}$
\begin{equation}
4\times 3\times \left(-\frac{1}{4}\right)\epsilon\Gamma^0\Gamma^{0\mu ij}\epsilon F_{9\mu}F_{ij}
= 8\tepsilon\Gamma^{09ij}\epsilon [\Phi_9,\Phi_j]\Phi_i,
\end{equation}

\item $\{M,N,P,Q\}=\{\mu,\nu,i,j\}$
\begin{equation}
\binom{4}{2}\left(-\frac{1}{4}\right)\epsilon\Gamma^0\Gamma^{\mu\nu ij}\epsilon F_{\mu\nu}F_{ij}
=-6\tepsilon\tGamma^0\tGamma^{ij}\Gamma^0\epsilon\Phi_iD_\nu \Phi_j.
\end{equation}
\end{enumerate}
Summarizing the above results of i)-vi), we obtain the gauge field part of $S^Q_{bos}$.
\begin{eqnarray}
S_{FF}&=& \frac{1}{2}BF_{MN}F^{MN}+\frac{1}{2}(2e^\Omega-B)F_{\mu\nu}*F^{\mu\nu}\nonumber\\
&& -4\tepsilon\Gamma^{09}\Gamma^{\mu\nu}\epsilon\Phi_9F_{\mu\nu}+8\tepsilon\Gamma^{09ij}\epsilon\Phi_i[\Phi_9,\Phi_j]-6\tepsilon\tGamma^\nu\tGamma^{ij}\Gamma^0\epsilon\Phi_iD_\nu \Phi_j.
\end{eqnarray}
%%%%%%%%%%%%%%%%%%%%
\subsubsection{Calculation of $S_{F\Phi}$}
The $S_{F\Phi}$ term
\begin{equation}
F_{MN}\Phi_A\tepsilon(\tGamma^A\tGamma^{NM}\Gamma^0+\tGamma^0\Gamma^A\Gamma^{MN})\epsilon
\end{equation}
is divided into three cases according to the index $A$.
The resulting non-zero contributions are the following cases. 
\begin{enumerate}
\def\theenumi{{\roman{enumi})}}
\item $A=0$
\begin{eqnarray}
S_{F\Phi} &=& F_{MN}\Phi_0\tepsilon (\tGamma^0\tGamma^{NM}-\Gamma^{MN})\epsilon\nonumber\\
&=& 2F_{0\mu}\Phi_0\partial^\mu B \s\text{(only $M=0$,$N=\mu$ case do not vanish.)}
\end{eqnarray}

\item $A=9$\\
The non-vanishing cases are $M=9,N\neq 9$ and $(M,N)=(\mu,\nu),(i,j)$. These results are
\begin{align}
&F_{MN}\Phi_9 \tepsilon (\tGamma^9\tGamma^{NM}\Gamma^0+\tGamma^0\Gamma^9\Gamma^{MN})\epsilon
 =F_{9N}\Phi_9 (-2\partial^N B),\nonumber\\
&F_{MN}\Phi_9 \tepsilon (\tGamma^9\tGamma^{NM}\Gamma^0+\tGamma^0\Gamma^9\Gamma^{MN})\epsilon
 =2F_{\mu\nu}\Phi_9\tepsilon\tGamma^0\Gamma^9\Gamma^{\mu\nu} + 2F_{ij}\Phi_9\tepsilon\tGamma^0\Gamma^9\Gamma^{ij}\epsilon,
\end{align}
respectively.

\item $A=i$\\
The non-zero cases are $M=j,N=\mu$ and $M=j,N=9$.
These results are
\begin{align}
&-4\Phi_iD_\mu\Phi_j\tepsilon\Gamma^{0ij\mu}\epsilon + 2\Phi^jD_\mu \Phi_j \partial^\mu B,\nonumber\\
&-4\Phi_i[\Phi_9,\Phi_j]\tepsilon\Gamma^{09ij}\epsilon,
\end{align}
respectively.\vspace{1cm}\\
\end{enumerate}

Gathering the results i)-iii), we obtain the result as follows:
\begin{eqnarray}
S_{F\Phi} &=& F_{0\mu}\Phi_0\partial^\mu B -2F_{9N}\Phi_9\partial^N B -2F_{j\mu}\Phi_j\partial^\mu B\nonumber\\
&& +2F_{\mu\nu}\Phi_9\tepsilon\Gamma^{09\mu\nu\epsilon} -6\Phi_i[\Phi_9,\Phi_j]\tepsilon\Gamma^{09ij}\epsilon -4\Phi_i(D_\mu \Phi_j)\tepsilon\Gamma^{0ij\mu} \epsilon.
\end{eqnarray}

Finally, we obtain the following deformation term.
\begin{eqnarray}
S^Q_{bos}&=&S_{FF}+S_{\Phi\Phi}+S_{KK}+S_{F\Phi}+S_{\Phi K}+S_{FK}\nonumber\\
		&=& \frac{1}{2}BF_{MN}F^{MN} +\frac{1}{2}(2e^{2\Omega}-B)F_{\mu\nu}*F^{\mu\nu}\nonumber\\
		&&	-4\tepsilon\Gamma^{09\mu\nu}\epsilon\Phi_9 F_{\mu\nu} +8\tepsilon\Gamma^{09ij}\epsilon[\Phi_9,\Phi_i]\nonumber\\
		&&	-4\Phi_A\Phi^A\tepsilon\tGamma^0\tepsilon +6\tepsilon\Gamma^{0\nu ij}\epsilon\Phi_iD_\nu \Phi_j
			-4\Phi_A\Phi^A\tepsilon\tGamma^0\tepsilon-BK_iK_i\nonumber\\
		&&	+2\Phi_0F_{0\mu}\partial^0 B -2\Phi_9F_{9N}\partial^NB -2\Phi_jF_{j\mu}\partial^\mu B\nonumber\\
		&&	+2\tepsilon\Gamma^{09\mu\nu}\epsilon\Phi_9F_{\mu\nu} -6\tepsilon\Gamma^{09ij}\epsilon\Phi_i[\Phi_9,\Phi_j] -4\tepsilon\Gamma^{0\mu ij}\epsilon\Phi_iD_\mu \Phi_j\nonumber\\
		&&	-4\Phi_0K_j\tepsilon\nu_j\\
		&=& \frac{1}{2}BF_{\mu\nu}F^{\mu\nu}+BD_\mu \Phi_A D^\mu \Phi^A +\frac{1}{2}B[\Phi_A,\Phi_B][\Phi^A,\Phi^B]+\frac{1}{2}(2e^{2\Omega}-B)F_{\mu\nu}*F^{\mu\nu}\nonumber\\
		&&	-2\tepsilon\Gamma^{09\mu\nu}\epsilon\Phi_9F_{\mu\nu}+2\tepsilon\Gamma^{0mij}\epsilon D_m\Phi_j\nonumber\\
		&&	-4\Phi_A\Phi^A\tepsilon\tGamma^0\tepsilon +2\Phi^AD_\mu \Phi_A\partial^\mu B -4K_i\Phi_0\nu_i\tepsilon -BK_iK_i.\label{SQbos1}
\end{eqnarray}
where we used in the last line the identity derived from $F_{A\mu}=-D_\mu\Phi_A$ and the relation
\begin{equation}
2\Phi_0F_{0\mu}\partial^0B - 2\Phi_9F_{9N}\partial^N B - 2\Phi_jF_{j\mu}\partial^\mu B=D_{\mu}(\Phi_A\Phi^A)\partial^\mu B.
\end{equation}

\subsection{Completing the square of $S^Q_{bos}$}
In preparation we calculate various spinor bilinears and their products.
The $\epsilon^{L,R}$ and $\epsilont^{L,R}$ defined in eq. \eqref{epsilonLR} are explicitly written as follows.
\begin{subequations}
\begin{align}
 &\epsilon^{L}=e^{\frac12 \Omega}\epsilon_s,\qquad
 \epsilon^{R}=e^{\frac12 \Omega}x^{a}\Gammat_{a}\epsilon_c,\\
 &\epsilont^{L}=Y^{ab}\Gammat_{ab}\epsilon_c+M\epsilon_c,\qquad
 \epsilont^{R}=e^{-\frac12 \Omega}\frac12 \del_{\nu}\Omega \delta^{\nu a}\Gamma_a\epsilon_s,
\end{align}
\end{subequations}
where $Y^{ab},M,\ (a,b=1,2,3,4)$ are defined as
\begin{subequations}
\begin{align}
& W^{ab}:=e^{-\frac12\Omega}\frac12 \del_{\nu}\Omega \delta^{\nu a} x^{b},\\
& Y^{ab}:=W^{[ab]}|_{\text{self-dual}},\\
& M:=\delta_{ab}W^{ab}+e^{-\frac12\Omega}.
\end{align}
\end{subequations}
The following formulas are quite useful.
\begin{equation}
\begin{aligned}
& \zeta \Gamma^{MNP} \zeta=0 \quad\text{ for any $\zeta$},\qquad
 \epsilon_c\epsilon_s=0,\qquad \epsilon_s \Gamma^{0} \epsilon_s=1,\qquad
\epsilon_c\Gammat^{0}\epsilon_c=-\frac{1}{t^2},\\
& \epsilon_c\Gamma_{12}\epsilon_s=
 \epsilon_c\Gamma_{34}\epsilon_s=1,\qquad
 \epsilon_c\Gamma_{ab}\epsilon_s=0 \quad\text{ if } [ab]\ne [12],[34]
\end{aligned} 
\end{equation}
The relation below with arbitrary spinors $\eta, \xi, \zeta$ can be derived from the triality identity.
\begin{align}
 \eta \Gamma_{M} \eta \xi \Gamma^{M} \zeta=-2\eta \Gamma_{M}\xi \eta \Gamma^{M}\zeta.
\end{align}
We also need the product of bilinears including $\nu_i$
\begin{align}
 \nu_i\epsilont \nu_i \epsilont
=-B\epsilont \Gammat^{0}\epsilont
 -\frac14 \del_{\mu}B \del^{\mu} B.\label{nuinui}
\end{align}
We define
\begin{align}
 f^{mij}:=\epsilont \Gamma^{0mij} \epsilon,\qquad m=1,2,3,4,9,\qquad i,j=5,6,7,8.
\end{align}
The square of this quantity is
\begin{eqnarray}
f^{mij}f^{mik}&=&\delta^{jk}\left\{
				-3(\tepsilon\epsilon)^2 +\frac{1}{2}\partial_\mu B\partial^\mu B -B\cdot\tepsilon\tGamma^0\tepsilon
				\right\},\\
\tepsilon\tGamma^0\tepsilon &=& -\frac{M^2+4Y^2}{t^2}-\frac{e^\Omega}{4}\partial_\mu\Omega\partial^\mu\Omega.
\end{eqnarray}
Other bilinears made of $\epsilon$ and $\epsilont$ are calculated as
\begin{align}
& \epsilont^{L}\epsilon^{L}=\epsilont^{R}\epsilon^{R}=-\frac4t e^{\frac12\Omega}Y_{12},\\
& \epsilont\epsilon=\epsilont^{L}\epsilon^{L}+\epsilont^{R}\epsilon^{R}=-\frac8t e^{\frac12\Omega}Y_{12},
\end{align}
\begin{align}
& \epsilon^{L}\Gamma^{0}\epsilon^{L}=e^{\Omega},\qquad \epsilon^{R}\Gamma^{0}\epsilon^{R}=\frac{x^2}{t^2}e^{\Omega}=B-e^{\Omega},\label{bilinears}\\
& \epsilont^{L}\Gammat^0 \epsilont^{L}=-\frac{1}{t^2}(M^2+4Y^{ab}Y^{ab}),\qquad
 \epsilont^{R}\Gammat^0 \epsilont^{R}=-\frac{1}{4}e^{\Omega}\del_{\mu}\Omega \del^{\mu}\Omega,\label{bilinears-t}\\
&\epsilont \Gammat^{0} \epsilont
 =-\left[
\frac{M^2+4Y^{ab}Y^{ab}}{t^2}+\frac14 e^{\Omega}\del_{\mu}\Omega \del^{\mu}\Omega
\right].
\end{align}
It is convenient to define the quantities $w_+^{ab}, w_-^{ab},\ (a,b=1,2,3,4)$ by
\begin{subequations}
\begin{align}
& w_+^{ab}:=-\frac{1}{e^{\Omega}}\epsilont^{L}\Gamma^{09ab}\epsilon^{L},\\
& w_-^{ab}:=-\frac{1}{B-e^{\Omega}}\epsilont^{R}\Gamma^{09ab}\epsilon^{R},
\end{align}
\end{subequations}
and the squares of them are calculated as
\begin{subequations}\label{w20}
\begin{align}
& e^{2\Omega}(w_+)^2=\epsilont^{L}\Gamma^{09ab}\epsilon^{L}\epsilont^{L}\Gamma^{09ab}\epsilon^{L}
=4(\epsilont^{L}\epsilon^{L})^2
 -8(\epsilont^{L}\Gamma^{09}\epsilon^{L})^2
 -4\epsilont^{L}\Gammat^{0}\epsilont^{L}\epsilon^{L}\Gamma^{0}\epsilon^{L},\\
& (B-e^{\Omega})^2(w_-)^2=\epsilont^{R}\Gamma^{09ab}\epsilon^{R}\epsilont^{R}\Gamma^{09ab}\epsilon^{R}
=4(\epsilont^{R}\epsilon^{R})^2
 -8(\epsilont^{R}\Gamma^{09}\epsilon^{R})^2
 -4\epsilont^{R}\Gammat^{0}\epsilont^{R}\epsilon^{R}\Gamma^{0}\epsilon^{R}.
\end{align}
\end{subequations}
We need to calculate the bilinears of the following in order to complete the calculation of the above
\begin{align}
 \epsilont^{L}\Gamma^{09}\epsilon^{L}
= -\epsilont^{R}\Gamma^{09}\epsilon^{R}
=\frac{4}{t}e^{\frac12\Omega}Y_{12}.
\label{09}
\end{align}
The results for $(w_+)^2, (w_-)^2$ are obtained by using eqs.\eqref{bilinears},\eqref{bilinears-t},\eqref{w20},\eqref{09},
\begin{subequations}\label{w2}
\begin{align}
& e^{\Omega}(w_+)^2=-\frac{64}{t^2}Y_{12}^2-4\epsilont^{L}\Gammat^{0}\epsilont^{L},\\
& (B-e^{\Omega})(w_{-})^2=-\frac{64}{x^2}Y_{12}^2-4\epsilont^{R}\Gammat^0 \epsilont^{R}.
\end{align}
\end{subequations}

We would like to rewrite the deformation term of the action, $S^Q_{bos}$, into the summation of positive definite terms. Let us put labels $I$ to $X$ to the terms in eq.\eqref{SQbos1} as 
\begin{eqnarray}
S^Q_{bos}&=& \frac{1}{2}BF_{\mu\nu}F^{\mu\nu} +BD_\mu\Phi_AD^\mu\Phi^A +\frac{1}{2}B[\Phi_A,\Phi_B][\Phi^A,\Phi^B] +\frac{1}{2}(2e^\Omega-B)F_{\mu\nu}*F^{\mu\nu}\nonumber\\
	&&	-2\tepsilon\Gamma^{09\mu\nu}\epsilon\Phi_9F_{\mu\nu} +2f^{mij}\Phi_iD_m\Phi_j\nonumber\\
	&&	-4\Phi_A\Phi^A\tepsilon\tGamma^0\tepsilon +2\Phi_AD_\mu \Phi^A\partial^\mu B -4K_i\Phi_0\nu_i\tepsilon -BK_iK_i\nonumber\\
	&=:& I+II+III+IV\nonumber\\
	&&	+V+VI\nonumber\\
	&&	+VII+VIII+IX+X.
\end{eqnarray}
We combine some of the terms and completing square.  First we manipulate the terms with gauge fields and $\Phi_9$.
\begin{align}
I+IV+V+VII|_{A=9}
&= \frac{1}{2}BF_{\mu\nu}F^{\mu\nu}+\frac{1}{2}(2e^\Omega-B)F_{\mu\nu}*F^{\mu\nu}
	-2\tepsilon\Gamma^{09\mu\nu}\epsilon\Phi_9F_{\mu\nu}-4\Phi_9^2\tepsilon\tGamma^0\tepsilon\nonumber\\
&= e^\Omega(F^+ +w^+\Phi_9)^2 +(B-e^\Omega)(F^- +w^-\Phi_9)^2\nonumber\\
&	-\underbrace{\{e^\Omega(w^+)^2+(B-e^\Omega)(w^-)^2+4\tepsilon\tGamma^0\tepsilon\}}
		_{=-(\frac{1}{x^2}+\frac{1}{t^2})64Y_{12}^2}\Phi_9^2,
\end{align}
where we use eqs.\eqref{w2}.

Second we take the terms with $K_i$ and $\Phi_0$
\begin{align}
IX+X+VII|_{A=0}
&= -BK_iK_i-4\Phi_0\nu_i\tepsilon K_i +4\tepsilon\tGamma^0\tepsilon\Phi_0^2\nonumber\\
&= -B\left(K_i+\frac{2}{B}\Phi_0\nu_i\tepsilon\right)^2 
	+\left\{\frac{4}{B}(\nu_i\tepsilon)(\nu_i\tepsilon)+4\tepsilon\tGamma^0\tepsilon\right\}\Phi_0^2.
\end{align}
This equation can be further simplified if we use eq.\eqref{nuinui}.

Third, the terms with $\Phi_A$ is rewritten as
\begin{align}
II+VIII &= BD_\mu\Phi_AD^\mu\Phi^A +2\Phi_AD_\mu\Phi^A\partial^\mu B\nonumber\\
&= B\left(D_\mu \Phi_A+\frac{\partial^\mu B}{B}\Phi_A\right)^2 -\frac{1}{B}\Phi_A\Phi^A\partial_\mu B\partial^\mu B.
\end{align}

Finally, there are the remaining terms $III+VI+VII|_{A=i}$.
Completing the square form of $S^Q_{bos}$ leaves 
\begin{eqnarray}
S^Q_{bos}
&=& e^\Omega(F_{+}+w_{+}\Phi_9)^2 + (B-e^\Omega)(F_{-}+w_{-}\Phi_9)^2 -B\left( K_i+\frac{2}{B}\Phi_0\nu_i\tepsilon \right)^2\nonumber\\
&&	+ B\Big\{ D_m\Phi_i+\frac{1}{B}(\Phi_i\partial_m B+f_{mji}\Phi^j)\Big\}^2 \nonumber\\
&&	-B\left( D_\mu\Phi_0 +\Phi_0\frac{\partial_\mu B}{B}\right)^2
	+B\left( D_\mu\Phi_9+\Phi_9\frac{\partial_\mu B}{B}\right)^2 
	\nonumber\\
&&	+B[\Phi_0,\Phi_i][\Phi^0,\Phi^i]+B[\Phi_0,\Phi_9][\Phi^0,\Phi^9]+\frac{1}{2}B[\Phi_i,\Phi_j][\Phi^i,\Phi^j]\nonumber\\
&&	+\left[\left(\frac{1}{t^2}+\frac{1}{x^2}\right)64Y_{12}^2-\frac{1}{B}\partial_\mu B\partial^\mu B\right]\Phi_9^2\nonumber\\
&&	
	-\frac{1}{B}\left\{-3(\tepsilon \epsilon)^2+\frac{3}{2}\partial_\mu B\partial^\mu B +3B\cdot \tepsilon\tGamma^0\tepsilon\right\}\Phi_i^2.
\end{eqnarray}

\providecommand{\href}[2]{#2}\begingroup\raggedright\endgroup

\end{document}